\documentclass[preprint]{aastex}
\slugcomment{To appear in ApJ, 2001 April 1}
\newcommand{\refer}{\reference}
\newcommand{\psiobl}{\psi_{\rm obl}}
\newcommand{\psipro}{\psi_{\rm pro}}
\begin{document} 
\title{Intrinsic Shapes of Molecular Cloud Cores}
\author{C. E. Jones\altaffilmark{1}, Shantanu Basu\altaffilmark{1}, and John
Dubinski\altaffilmark{2}}
\email{cjones@io.astro.uwo.ca}
\email{basu@astro.uwo.ca}
\email{dubinski@cita.utoronto.ca}
\altaffiltext{1}{Department of Physics and Astronomy, University of
Western Ontario, London, Ontario, Canada N6A~3K7.}

\altaffiltext{2}{Department of Astronomy, University of Toronto, 60 Saint George Street, Toronto, Ontario, Canada M5S~3H8.}

\begin{abstract}
We conduct an analysis of the shapes of molecular cloud cores using 
recently compiled catalogs of observed axis ratios of individual cores
mapped in ammonia or through optical selection. We apply
both analytical and statistical techniques to de-project the observed axis 
ratios in order to determine the true distribution of cloud core shapes. 
We find that neither pure oblate nor pure prolate cores can account for
the observed distribution of core shapes.  Intrinsically triaxial cores
produce distributions which agree with observations.  The best fit
triaxial distribution contains cores which are more nearly oblate
than prolate.  
\end{abstract}

\keywords{ISM: clouds --- ISM: globules --- ISM: structure ---stars: formation }

\section{Introduction}

Molecular clouds are the sites of star formation in our Galaxy.
Star formation occurs within dense cores in these clouds
(Myers \& Benson 1983; Benson \& Myers 1989), as evidenced by their
correlation with young stars (Beichman et al. 1986).
While difficult to de-project, information about the intrinsic
shapes of cores can yield insight into which physical processes
control their evolution and thereby govern star formation. 

Observational estimates of energy densities imply that many cores 
have magnetic, turbulent, and gravitational energies which are compatible
with the cores being in virial equilibrium 
(Myers \& Goodman 1988). Theoretical models of cloud equilibria
routinely invoke axisymmetry, so that magnetically supported (Mouschovias 1976)
or rotationally supported (e.g., Kiguchi et al. 1987)
clouds, which are flattened in a preferred direction, assume an oblate shape. 
Such equilibria are analogous to axisymmetric oblate equilibrium 
models applied successfully to understand the structure of stars, planets,
and accretion disks.

By analyzing the apparent shapes of 16 cores, Myers et al. (1991) concluded 
that the mean apparent axis ratio $\langle p \rangle$ $\approx$ 0.5 was 
consistent with 
oblate objects of intrinsic axis ratio $q \approx $ 0.1 - 0.3 and
prolate objects with $q \approx $ 0.5. Based on a preference for less 
flattened objects and the suggestive elongation of some cores along 
elongated parent clouds, Myers et al.  (1991) concluded that cores were 
more likely to be prolate axisymmetric
objects than oblate axisymmetric objects. Ryden (1996) looked at the
distribution of core shapes from several catalogs, including the 
Benson \& Myers (1989) catalog of ammonia cores
(41 objects) and the Clemens \& Barvainis (1988) catalog of Bok 
globules. She concluded that the shape distribution of dense cores was
more consistent with prolate rather than oblate objects, assuming that the
cores were axisymmetric. The projected shapes of Bok globules were found to 
be consistent with either hypothesis, with oblate objects requiring an 
intrinsic mean axis ratio of 0.3, and prolate objects requiring a mean 
ratio of 0.5. Li \& Shu (1996) have also
pointed out that oblate objects with intrinsic axis ratio $\approx$ 0.3
can explain the observed $\langle p \rangle$ $\approx$ 0.5. 
Ciolek \& Basu (2000) present a specific case of how an oblate cloud with
such an axis ratio can fit the observations of a single core.

The apparent observational bias toward prolate axisymmetric objects 
has led to some models of equilibrium prolate objects (Tomisaka 1991;
Fiege \& Pudritz 2000) which require a dominant role for toroidal 
(rather than poloidal) magnetic fields, in order to maintain an equilibrium
along the long axis. 
In practice, it is difficult to generate a model 
that can sustain a prolate geometry, as collapse along a poloidal
field (parallel to the the long axis) results in oblate 
objects (Nakamura et al. 1995). Therefore, equilibrium prolate objects
supported by toroidal field pressure require a constant source of
significant magnetic helicity.

It has also been suggested that the apparently prolate shapes
are an indication that cores are not in equilibrium (Fleck 1992). In fact,
since many cores appear to be self-gravitating, it is natural to assume
that gravitational motions (likely softened due to magnetic and/or turbulent 
support) are an important component in determining their evolution.
An important feature of gravity is that it tends to enhance anisotropies in
collapsing bodies (Mestel 1965; Lin, Mestel, \& Shu 1965), so that 
unstable objects first collapse along one dimension, then break up into 
elongated objects within the sheet. While these objects may eventually 
break up into 
near-spherical fragments, this scenario implies that most objects have no 
intrinsic symmetry; thus a more general configuration 
is a triaxial, rather than an axisymmetric body. 
While studies of individual collapsing fragments often assume axisymmetry
(e.g., Basu \& Mouschovias 1994),
as a consequence of equilibrium (or near-equilibrium) initial states, 
it is likely that such objects will not remain axisymmetric if the numerical 
restriction of axisymmetry is relaxed (e.g., see Nakamura \& Hanawa 1997).
Additionally, large scale turbulence in molecular clouds will also likely cause
the initial states for collapse to deviate from perfect axisymmetry.

New surveys (Lee \& Myers 1999; Jijina, Myers, \& Adams 1999) have for the 
first time cataloged core properties for several hundred objects, allowing
a more meaningful statistical analysis of core properties.
Jijina et al. (1999) compiled a catalog of 264 dense cores from NH$_3$ 
observations, and Lee \& Myers (1999) compiled a catalog of 406 dense cores 
from contour maps of optical extinction.
With such data sets, the {\it distribution} of apparent projected core axis 
ratio $p$ (not just the mean value $\langle p \rangle$), can be used to 
constrain the intrinsic core shapes, similar to what has been done in
the field 
of galaxy studies (see Lambas, Maddox \& Loveday 1992; also discussion 
in Binney \& Merrifield 1998)

In this paper, we conduct an analysis of the observed core shape 
distributions in order to determine their intrinsic shapes. We check
the possibility that cores may be axisymmetric
(prolate or oblate) objects in \S\ 3, and consider the more general 
possibility of triaxial objects in \S\ 4. A discussion and summary are given
in \S\S\ 5 and 6, respectively.

\section{Theoretical Background}

Any given oblate, prolate, or triaxial body, when viewed in 
projection, yields a distribution of observed projected axis ratios 
$\phi(p)$ arising from the different possible viewing angles. 
More generally, one often wishes to determine the distribution of 
intrinsic axis ratios based on the observed $\phi\left(p\right)$.
Analytical methods have been developed to determine the intrinsic
shapes of elliptical galaxies and globular clusters (for example, 
Sandage, Freeman \& Stokes 1970; Binney
1978; Fall \& Frenk 1983) from the observed $\phi(p)$,
assuming intrinsically axisymmetric objects. We return to the axisymmetric case
in \S\ 3.

\begin{figure}
\epsscale{1.0}
\plotone{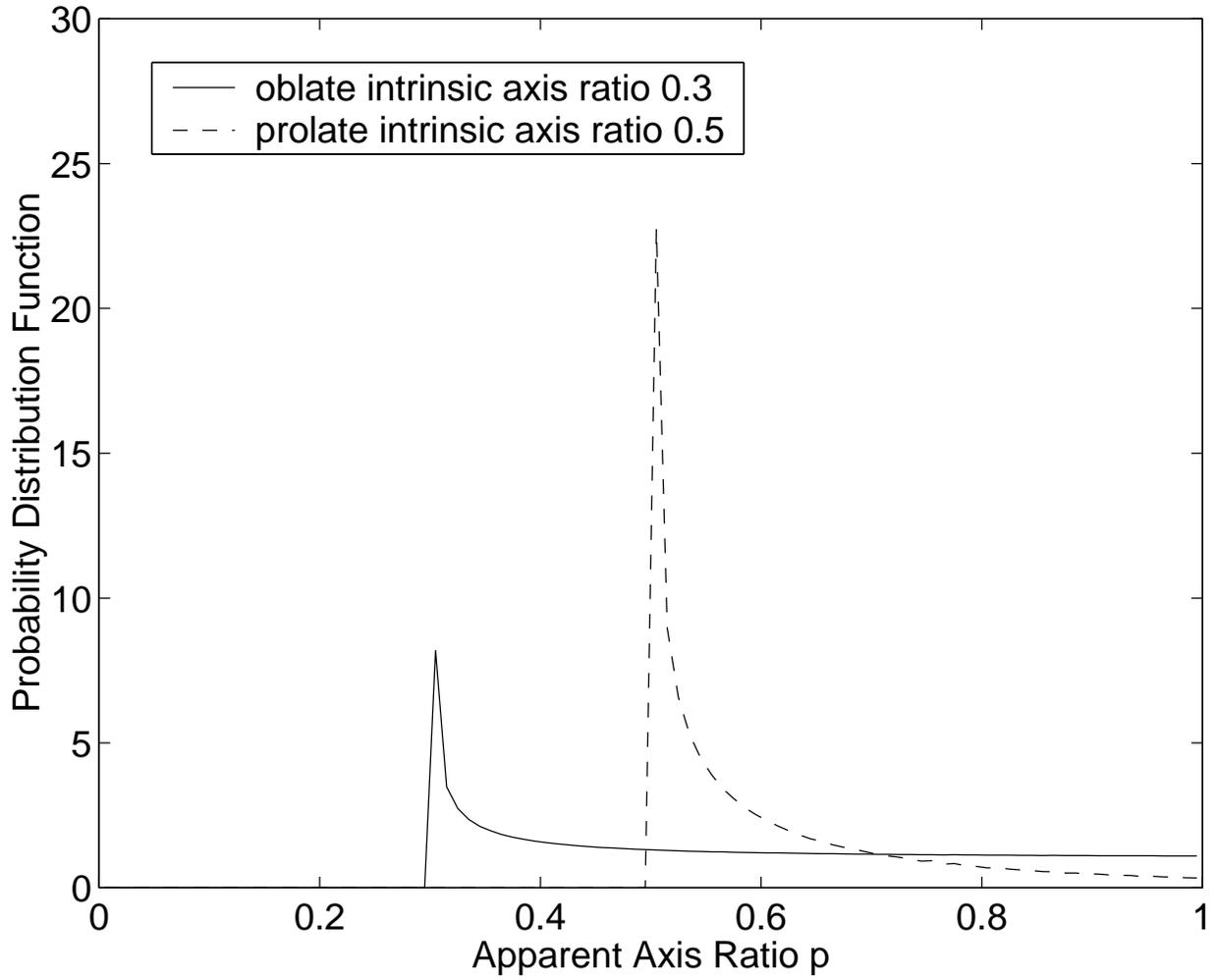}
\caption{Probability density of observing the projected axis ratio $p$
assuming either a pure oblate intrinsic shape with an axis ratio of 0.3 or pure
prolate intrinsic shape with an axis ratio of 0.5.}
\label{binneyobandpro}
\end{figure}

In general, a triaxial ellipsoid can be described by the equation
\begin{equation}
x^2 + \frac{y^2}{\zeta^2} + \frac{z^2}{\xi^2} = a^2,
\end{equation}
where $a$ is a constant and $1 \ge \zeta \ge \xi$.  
The geometrical analysis of Stark (1977) and Binney (1985) shows that such
a body, when viewed in projection, has elliptical contours. Following 
Binney (1985), the projection of a triaxial body when viewed from an
observing angle $(\theta, \phi)$ (where the
angles are defined on an imaginary viewing sphere and have their usual
meaning in a spherical coordinate system) is found using the quantities
\begin{equation}
A \equiv \frac{{\cos}^2 \theta}{\xi^2}\left({\sin}^2\phi +
\frac{{\cos}^2\phi}{\zeta^2}\right) + \frac{{\sin}^2\theta}{{\zeta}^2},
\end{equation}
\begin{equation}
B \equiv \cos \theta \, \sin2\phi \left(1 - \frac{1}{\zeta^2}\right) \frac{1}{\xi^2},
\end{equation}
and
\begin{equation}
C \equiv \left(\frac{{\sin}^2 \phi}{\zeta^2} + {\cos}^2\phi \right)
\frac{1}{\xi^2}.
\end{equation}
The apparent axis ratio in projection then equals
\begin{equation}
p = {\left(\frac{A + C - D}{A + C + D}\right)}^{1/2},
\label{binneyform}
\end{equation}
where $D \equiv \sqrt{{\left(A - C \right)}^2 + B^2}$. 
Using these equations, one can construct probability distributions
for the projected axis ratio, assuming a large
number of randomly distributed viewing angles.
If we assume that cores are either
oblate or prolate axisymmetric objects, then the above equations
simplify further since $\zeta = \xi$ for the prolate case, and
$\zeta = 1$ for the oblate case.
Figure~\ref{binneyobandpro} shows an example of the distributions
generated from these equations for axisymmetric cores.  
The oblate object is assumed to have an axis ratio of 0.3 and
the prolate object an axis ratio of 0.5.  Similar figures are
presented in Binney \& Merrifield (1998). 
These distributions peak at a value near the intrinsic axis ratio and then fall off
rapidly.  The oblate distribution (solid line) levels off to a
near-constant value whereas the 
prolate distribution (dashed line) continues to decrease.
The prolate distribution is also much more strongly peaked near its
maximum. Both phenomena are related to the intuitive result that prolate 
(i.e., filamentary) objects appear elongated with an apparent axis ratio 
close to the true value when seen from most viewing angles. There is only a 
very rare chance of viewing them nearly along their long axis, from which
they appear circular. Interestingly, both curves
in Figure~\ref{binneyobandpro} have the same mean $\langle p \rangle$ $\approx
0.6$.  This is evidence that simply fitting a model to the observed
$\langle p \rangle$ may not yield reliable information about the 
intrinsic distribution of core shapes.    

\begin{figure}
\plotone{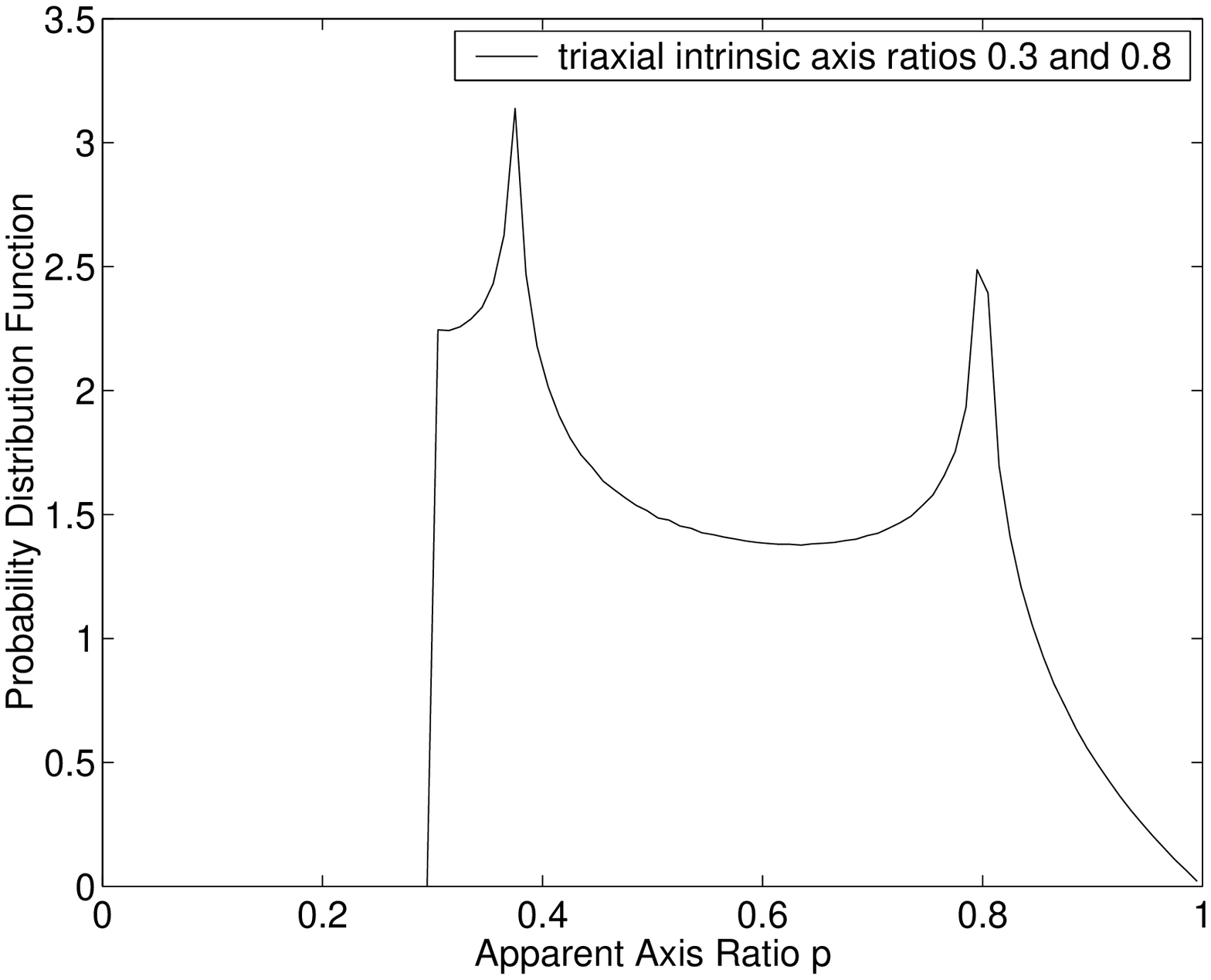}
\caption{Probability density of observing the projected axis ratio $p$
assuming a triaxial shape with intrinsic axis ratios of 0.3
and 0.8.}
\label{binneytri}
\end{figure}

What would the observed distribution look like if cores were triaxial
with particular values for the two axis ratios, $\zeta =  b/a$ and
$\xi = c/a$?
Figure~\ref{binneytri} shows the resulting distribution assuming a
triaxial object with axis ratios of 0.3 and 0.8. Notice that the 
distribution, besides
having peaks near 0.3 and 0.8, falls to zero at $p=1$, i.e., the body
never appears circular in projection.
Hence, the differences in shape of the predicted
probability distribution, especially in the
vicinity of $p=1$, act as an important discriminant in determining which 
intrinsic shapes best produce an observed distribution. 

The figures shown here were obtained by using a Monte Carlo program which
calculates the expected observed distribution of axis ratios for an
assumed intrinsic triaxial body (or distribution of triaxial bodies).
The triaxial bodies can be prolate or oblate in special cases.
The analytical expression (\ref{binneyform}) for $p$ is evaluated for a 
prescribed number
of randomly placed viewing angles ($\theta, \phi$). This program is a
modified version of one used by Dubinski \& 
Carlberg (1991), and is
used extensively to investigate triaxial shapes of cores in \S\ 4.

\section{Analytical Application to Ammonia Cores}

\begin{figure}
\plotone{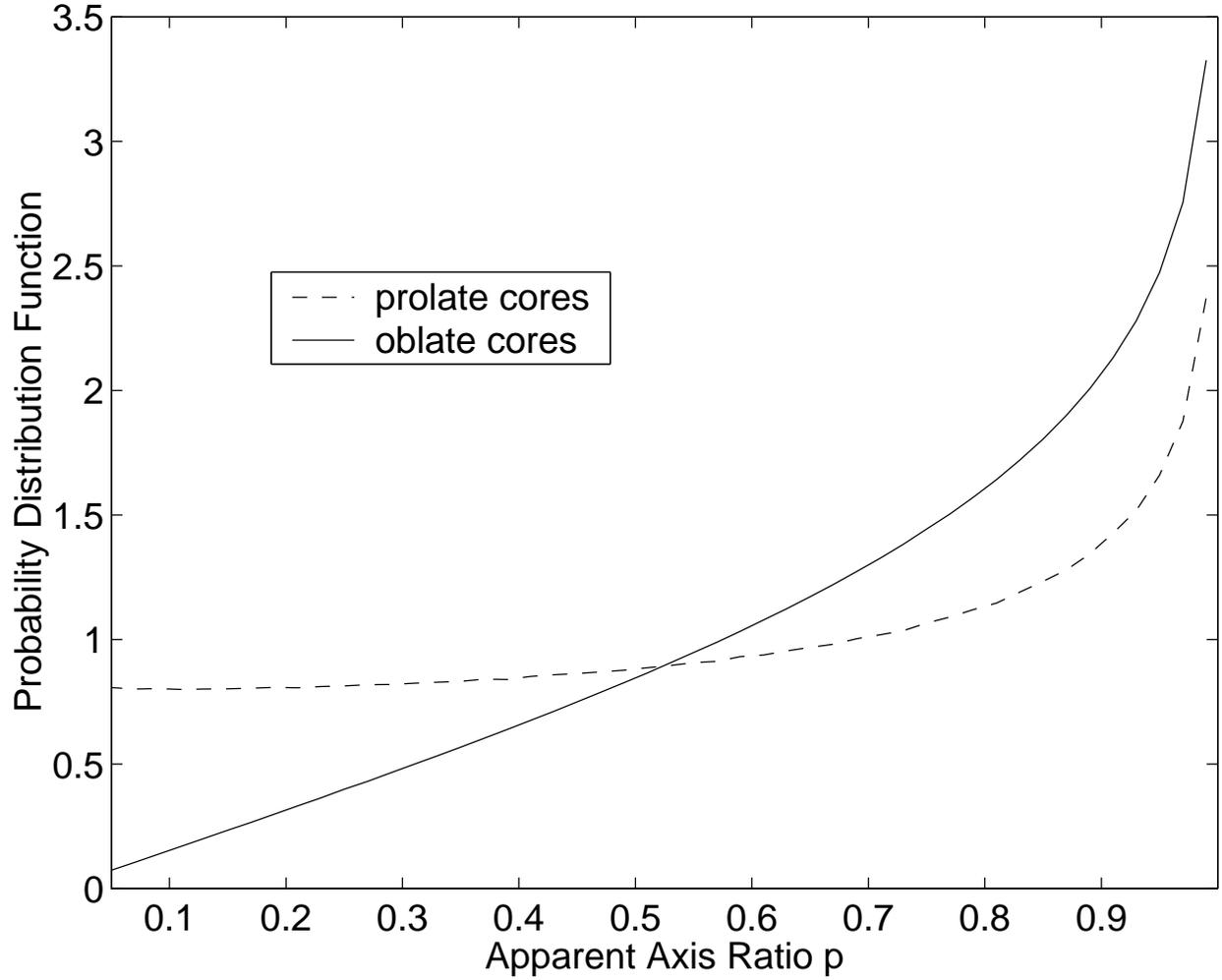}
\caption{Probability density of observing the projected axis ratio $p$
assuming either an axisymmetric oblate (solid line) or prolate (dashed
line) intrinsic shape
and a uniform distribution of axis ratios.}
\label{uniform}
\end{figure}

In this section, we consider the axisymmetric assumption of either 
prolate or oblate cores. Now, if we assume axisymmetric cores
with any intrinsic axis ratio $q$ equally likely, what is the expected
distribution? Figure~\ref{uniform} shows the resulting 
distribution for intrinsic axis
ratios distributed uniformly over the parameter space.  This figure
shows that for both axisymmetric shapes, the distribution
peaks at large values of the projected axis ratio $p$; in other words, the 
projected probability distributions of objects with high $q$ (which have 
a strong peak at $p \gtrsim q$) dominate the overall distribution.
This is clearly different than the actual observed 
distributions; (see Fig. 4 for 
example).  Therefore, if these
objects are axisymmetric, certain values of the axis ratios must be
favored.   

Equations have been constructed which relate the 
observed distribution $\phi(p)$ to the distribution of intrinsic
axis ratios $\psi(q)$, assuming either axisymmetric oblate or prolate objects 
(Fall \& Frenk 1983). For pure oblate shapes,
\begin{equation}
\phi\left(p\right)=p\int^{p}_{0}dq\left(1-q^2\right)^{-\frac{1}{2}}\left(p^2-q^2\right)^{-\frac{1}{2}}\psi\left(q\right),
\end{equation}
and for pure prolate shapes,
\begin{equation}
\phi\left(p\right)=p^{-2}\int^{p}_{0}dq\,q^2\left(1-q^2\right)^{-\frac{1}{2}}\left(p^2-q^2\right)^{-\frac{1}{2}}\psi\left(q\right).
\end{equation}
Lambas et al. (1992) use a polynomial fit to the observed data, of the
form
\begin{equation}
\phi\left(p\right)=c_{1}p + c_{3}p^3 + c_{5}p^5.
\end{equation}
They subsequently invert equations (6) and (7) to obtain
\begin{equation}
\psiobl\left(q\right)=\frac{2}{\pi}\left(1-q^2\right)^{1/2}\left(c_1 + 2c_3 \, q^2 + \frac{8}{3}c_5 \, q^4\right)
\end{equation}
and 
\begin{equation}
\psipro\left(q\right)=q^{-2}\left(1-q^2\right)^{1/2}\left(\frac{3}{2}c_1 \, q^3 + \frac{15}{8}c_3 \, q^5 + \frac{35}{16}c_5 \, q^7\right),
\end{equation}
using the methods of Fall \& Frenk (1983).
Their results show that both $\psiobl$ and $\psipro$
become negative for
values of $q$ greater than 0.9.  They conclude that neither pure
prolate or pure oblate models adequately describe the intrinsic shape
of elliptical galaxies. A similar analysis by Ryden (1996), using a
non-parametric kernel method, showed that the observed shape distribution of
dense cores available in several catalogs were more consistent with 
intrinsically prolate (rather than oblate) spheroids, although the cores
mapped in ammonia were inconsistent with both the prolate and oblate
hypothesis.

\begin{figure}[h]
\plotone{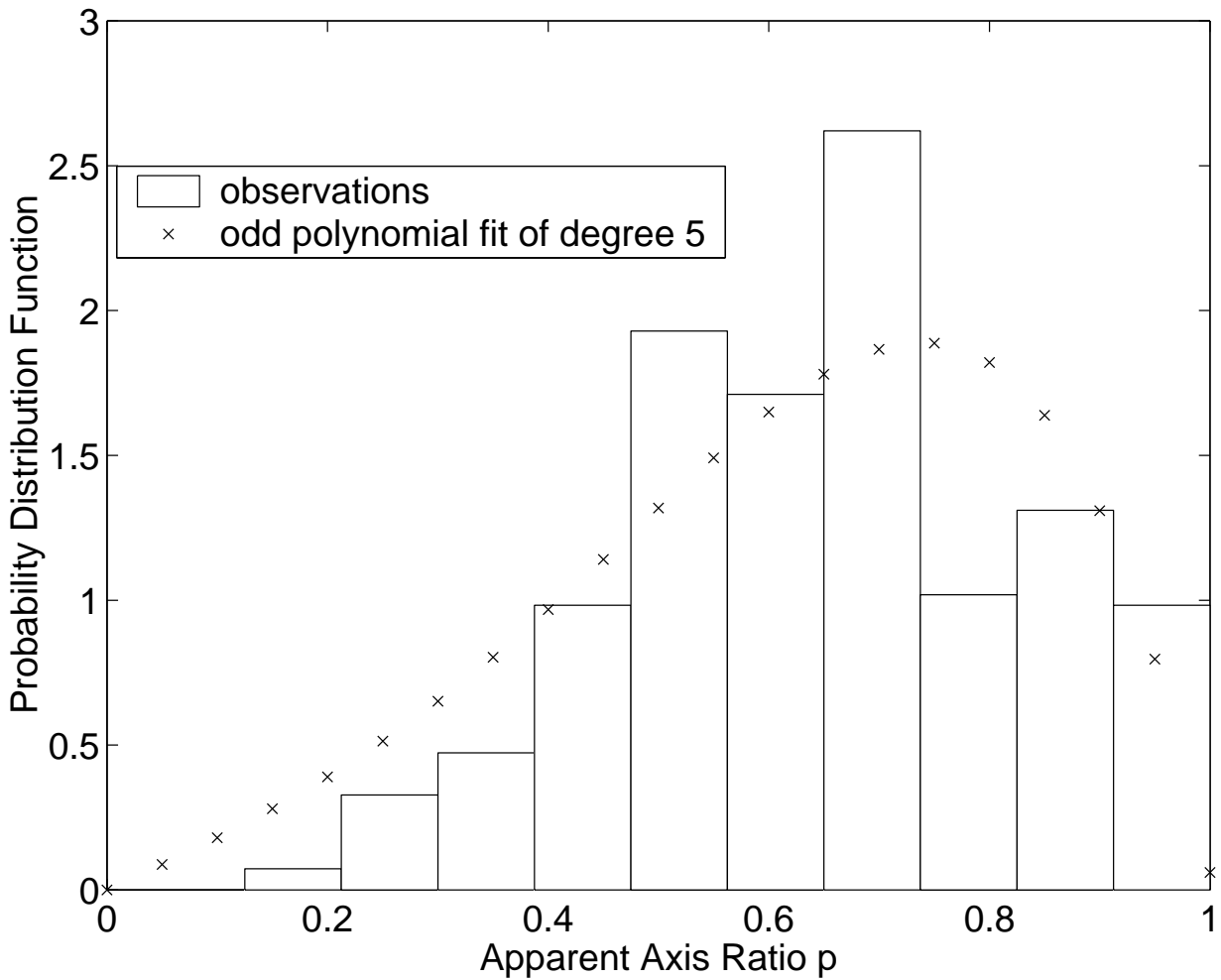}
\caption{A comparison of the binned data set (a normalized histogram) of projected axis ratios
for the cores mapped in ammonia (Jijina et al. 1999) with the odd polynomial fit.}
\label{jijdatafit}
\end{figure}

\begin{figure}[h]
\plotone{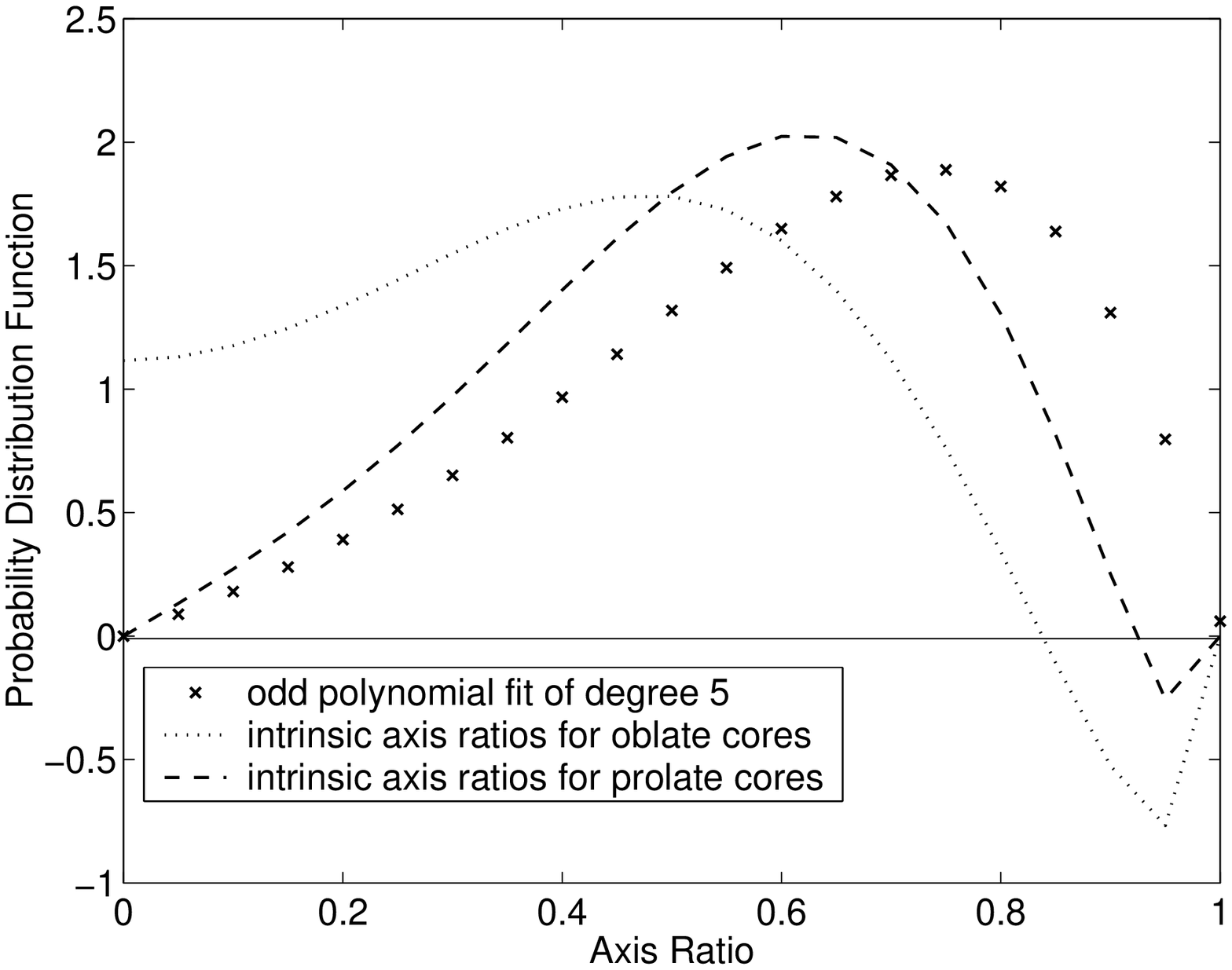}
\caption{The analytically derived intrinsic axis ratio distributions 
$\psiobl$ (dotted line) and $\psipro$ (dashed line) 
for the Jijina et al. (1999) data. The polynomial fit to the
observed axis ratio distribution (crosses) is also shown.}
\label{jijfits}
\end{figure}

Here, we also apply an analytical inversion method, assuming axisymmetry,
to the recent survey by Jijina et al. (1999) of dense cores mapped in ammonia.
Based on the size and estimated uncertainties of the Jijina et al. data set, 
a histogram of the observed axis ratios $p$ is 
created with 10 bins to sample the data adequately.  The observations
are then fit with polynomials of various degrees in order to
find the best fit.   Following the approach by Lambas et al. (1992),
we find that an odd polynomial of degree 5 represents a good fit to the
data, with $c_1=1.75, c_3=5.28$, and $c_5=-6.97$. Figure~\ref{jijdatafit} 
shows the histogram of the original data and the polynomial fit.
Equations (9) and (10) are then evaluated using this polynomial fit.
The resulting distributions $\psiobl$ and $\psipro$ 
are displayed graphically in Figure~\ref{jijfits}.
For both the assumptions of oblate cores and prolate
cores, the predicted intrinsic distribution of axis ratios becomes
negative for large values of $q$. Clearly this resulting distribution is
unphysical.  Either the initial assumption of axisymmetry is
incorrect, or errors in one or more of the binning, polynomial fits, 
or observations have caused this unphysical result.  Also, for this 
analysis to be applicable, the cores must be randomly 
oriented in space. For specific star forming regions
there may be some preferred orientation of the cores; however, this
catalog samples many areas of the sky and is composed of hundreds of
data points, so we assume that the cores are randomly oriented.

To estimate the effect of observational, binning, 
and polynomial fitting errors, we apply a technique known as
boot-strapping, which requires removing and adding 
data to various bins in the data set.  We are most interested in 
changes in the values of the projected axis ratios $p \gtrsim 0.8$, which
determine whether or not the values of the intrinsic axis ratio $q$ can
become positive. Removing data in these bins produces oblate and prolate
fits which are more negative.  Adding data at large $p$ increases the
probability distribution function for large $q$; however, we find that adding 
up to 20\% more data values of $p$ in the range from 0.8 to 0.9 and 0.9 to 1.0 
produces intrinsic distributions which still remain negative for large 
values of $q$.   
We conclude that neither pure oblate nor pure prolate
distributions can reproduce the observed axis ratios.

%




\section{Analysis of Triaxial Cores}

\begin{figure}
\plotone{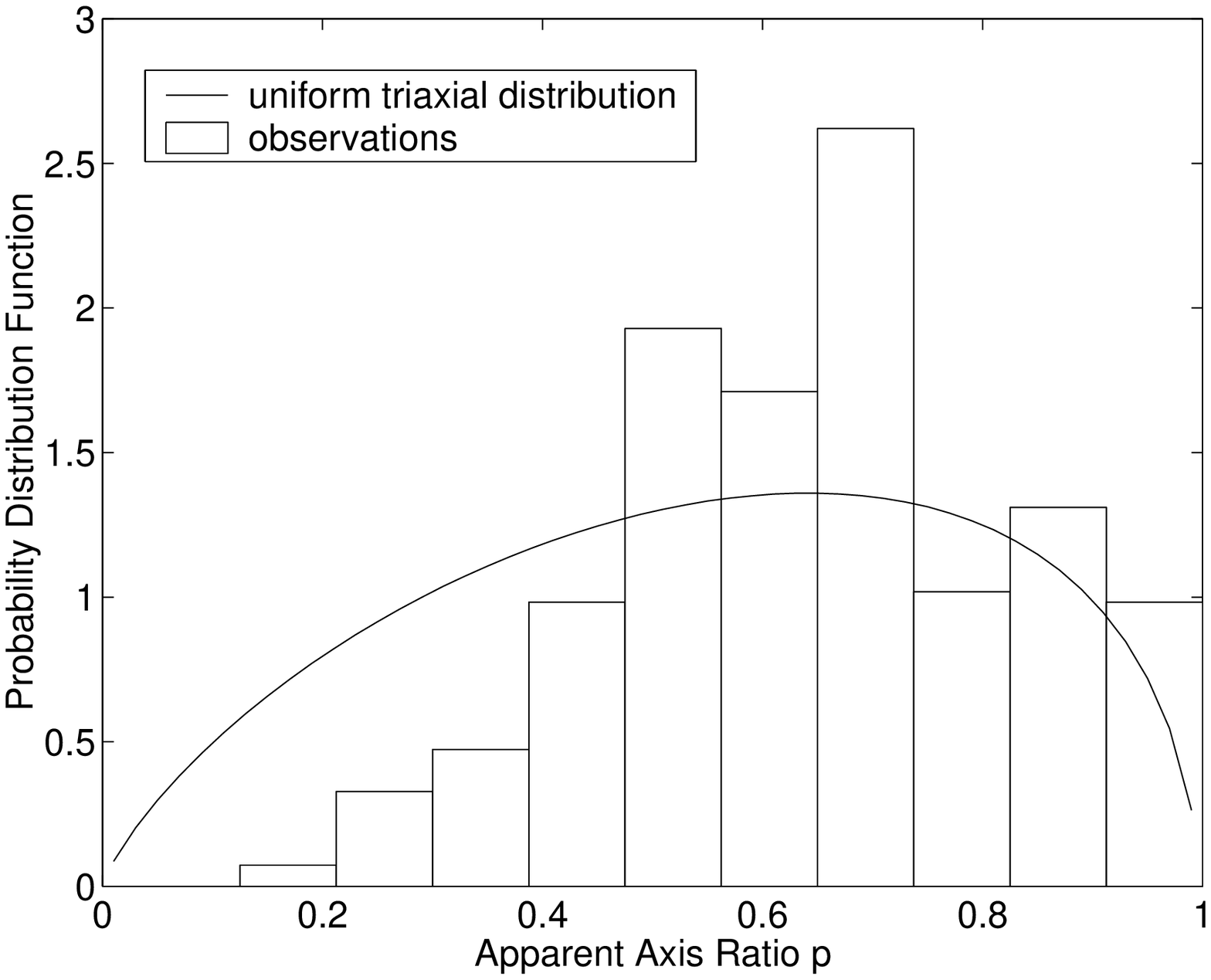}
\caption{Probability density of observing the projected axis ratio $p$
assuming a triaxial shape 
for a uniform distribution of axis ratios.  Also shown is the binned data set
(Jijina et al. 1999)
of the cores mapped in ammonia for comparison.}
\label{uniformtri}
\end{figure}

Since we reject the possibility that cores
are axisymmetric in \S\ 3, we now consider a more general
shape for cores.
If we assume, for example, that these objects are triaxial
with any intrinsic axis ratio equally likely, what is the expected
observed distribution? Figure~\ref{uniformtri} shows the resulting 
distribution from our Monte Carlo program, assuming triaxial cores with axis
ratios $\zeta$ and $\xi$ distributed uniformly in the range [0,1].
(Also shown is a normalized histogram of the observed ammonia cores.)
This distribution turns down towards $p=1$,
as all distributions for individual triaxial bodies do, and is therefore
a better fit to observations than a uniform
distribution of axisymmetric objects (see Fig. \ref{uniform}). 
However, it is still
not a good fit to the observations. The distribution greatly overestimates 
the number of objects with axis ratios $\leq 0.4$, does not fit the
peak of the observed distribution near 0.6, and underestimates the
number of objects with axis ratios $\geq 0.9$.  Therefore, 
if cores are triaxial in shape, certain values of axis ratios must be
preferred.   

We conduct an investigation in order to determine the
most likely values of the axis ratios $\zeta$ and $\xi$ for a distribution
of triaxial shaped cores. We assign a Gaussian distribution 
for each axis ratio $\zeta$ and $\xi$, with a mean in the range $[0,1]$, 
and standard deviation $\sigma$ typically equal to 0.1, consistent
with our use of 10 bins to sample the data. We did test a range of 
$\sigma$'s from 
0.05 to 0.2, and present the best fits for $\sigma = 0.1$ in this paper.
We find that our conclusions do not change significantly within this range 
of $\sigma$ values. The drawback to using relatively large $\sigma \gtrsim 0.2$ 
is that a relatively large fraction of the Gaussian distribution falls
outside the allowed range $[0,1]$. 
We tested several different ways to deal with these values:
(1) we set values less than zero to zero and values greater
than one to one; (2) we removed all numbers outside of the
range of zero to one; (3) we rejected numbers which
fell outside of the range of zero to one and repeatedly generated a
new random number until one between zero and one was obtained.
None of these methods is completely
satisfactory since Gaussian distributions modified by these methods
clearly have different means and standard deviations than originally
specified.  We believe that it is not very meaningful to test Gaussian
distributions with larger values of $\sigma$ than 0.2 especially near
the end-points of our parameter space.  For example, a Gaussian
distribution centered at 0.8 with a $\sigma$ of 0.2 has $\approx$ 16\%
of the data $>1$. If one simply truncates the values
$>1$, the new resultant distribution is centered at 0.7 with a
standard deviation of 0.16.  For similar reasons, we limit the $\sigma
= 0.1$ analysis to the range of axis ratios [0.1,0.9].

\begin{figure}[h]
\plotone{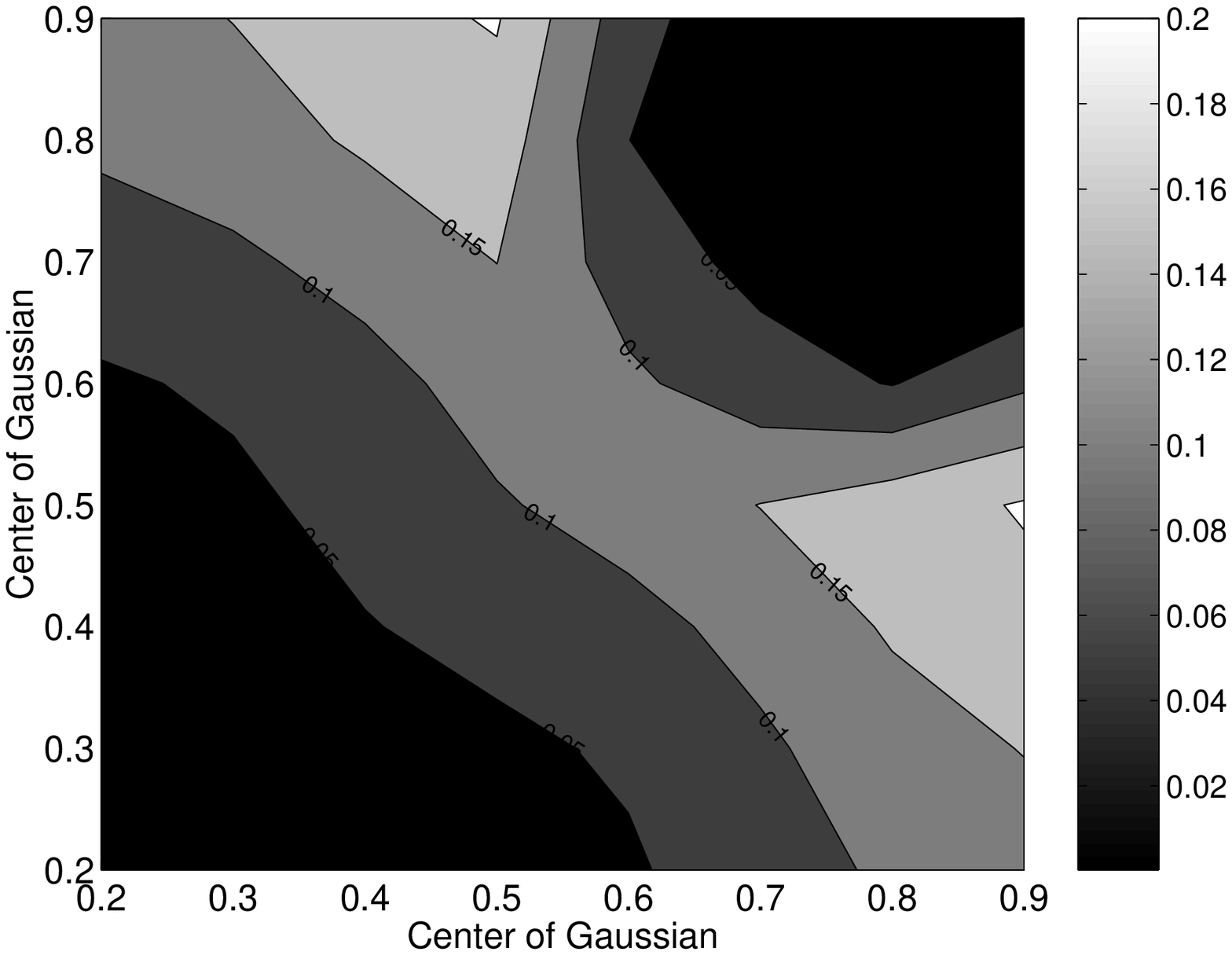}
\caption{A 2-dimensional plot of inverse $\chi^2$ values for triaxial
core shape models for the Jijina et al. (1999) data.  Note the
symmetry of the figure about the line along which the two centers of the
axis distributions are equal.  For any point in the figure, the
smaller axis ratio corresponds to $\xi_0$ and the larger to $\zeta_0$.}
\label{trijij}
\end{figure}

\begin{figure}[h]
\plotone{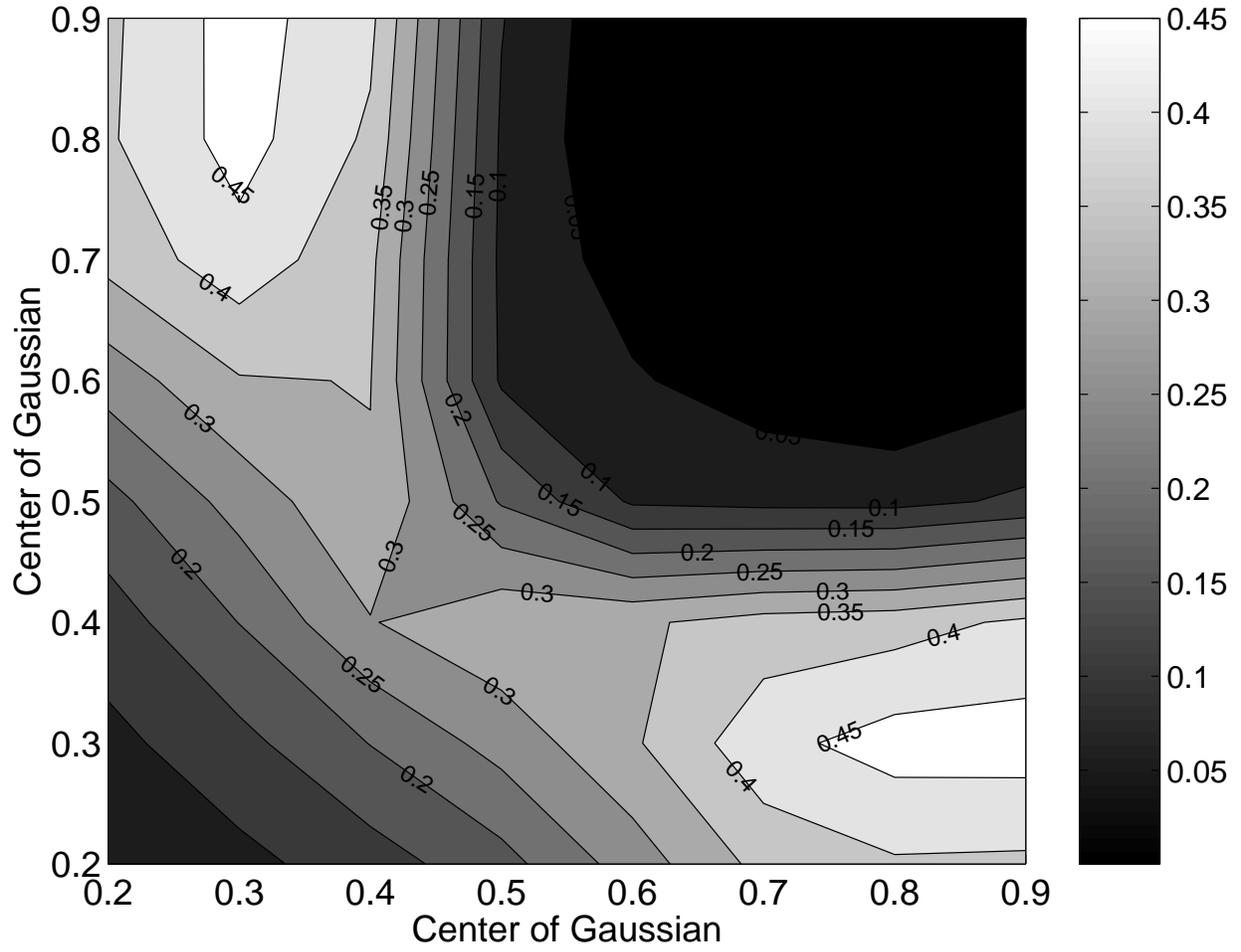}
\caption{A 2-dimensional plot of inverse $\chi^2$ values for triaxial
core shape models for the Lee \& Myers (1999) data.}
\label{trilee}
\end{figure}

In order to find the best fit intrinsic distribution of triaxial bodies, 
distributions of axis ratios with peak values $\xi_0$ and $\zeta_0$ 
(for a given $\sigma$) are
input into the Monte Carlo program described in \S\ 2. We typically 
employ at least $10^4$ viewing angles to calculate the projected distribution
for each pair of axis ratios, and at least $10^4$ 
axis ratios for each Gaussian distribution. This is more than sufficient for 
comparison with data sets sampled in only 10 bins. The program
produces the expected observed distribution which results from the assumed
intrinsic distributions. We compare this output with the observed data sets
of both Jijina et al. (1999) and Lee \& Myers (1999).  The best fits 
are determined by comparing with the data bins and calculating the 
$\chi^2$ value.
Figure~\ref{trijij} and Figure~\ref{trilee} show 2-dimensional plots
of the inverse $\chi^2$ values for the Jijina et al. data and the Lee
\& Myers data, respectively.  
Both data sets are fit best by a
triaxial distribution with $\xi_0$ just below the observed mean 
$\langle p \rangle$ and $\zeta_0$ close to unity.
The Jijina et al. data set is best fit by distributions with 
axis ratios $(\xi_0,\zeta_0) = (0.5,0.9)$, and the Lee \& Myers data 
set is best fit by values $(\xi_0,\zeta_0) = (0.3,0.9)$.
For comparison, Jijina et al. (1999) calculate a mean
projected aspect ratio $\langle p \rangle^{-1} \approx 1.5$,
or equivalently $\langle p \rangle \approx .67$.
Lee \& Myers state a value of $\langle p \rangle^{-1} \approx 2.4$,
or equivalently $\langle p \rangle \approx .42$ for their data set.
(For the Lee \& Myers data set, we calculate
a mean projected aspect ratio of 2.0 directly from their data set.
It is unclear what the reason
for this discrepancy is.)  Nevertheless, our main conclusions are clear:
(1) a triaxial distribution rather than an axisymmetric distribution
can best produce the observed distribution of shapes; (2) the
most likely distribution of triaxial shapes has more near-oblate objects
than near-prolate ones, since $\zeta_0$ is always close to 1. 
(Formally, one may say that the cores are more nearly oblate if 
$\zeta_0 > \onehalf [1 + \xi_0]$.)

The differences in the best fit axis ratios ($\xi_0,\zeta_0$) for the two
data sets may simply reflect differences in the intrinsic
shape of these cores when mapped in ammonia lines compared to
optical observations. Lee \& Myers (1999) state that the optical maps
trace regions of mean density $n \simeq (6-8) \times 10^3$ cm$^{-3}$, which is 
slightly lower than the $n \gtrsim 10^4$ cm$^{-3}$ sampled in ammonia 
maps. Hence, the greater mean elongation (lower $\langle p \rangle$)
of the optically selected cores, which results in a lower best-fit 
$\xi_0$, may be related to the theoretical result that outer density
contours of core models are often more elongated than inner ones
(e.g., Fiedler \& Mouschovias 1993). This is due to the isotropic thermal
pressure support being relatively more important on smaller scales.

In order to test the robustness of our result, we remove 20\% of the data in
each sample randomly and recalculate the
$\chi^2$ values.  This is repeated
10 times for each data set.  Each such simulation with the Jijina et al.
(1999) data set continues to have a best fit at mean axis ratios 0.5 and 0.9.
The Lee \& Myers
data has a best fit with one mean axis ratio 0.3 in all cases, and the
other in the range 0.7 to 0.9, with a mean of 0.83.

\begin{figure}[h]
\plotone{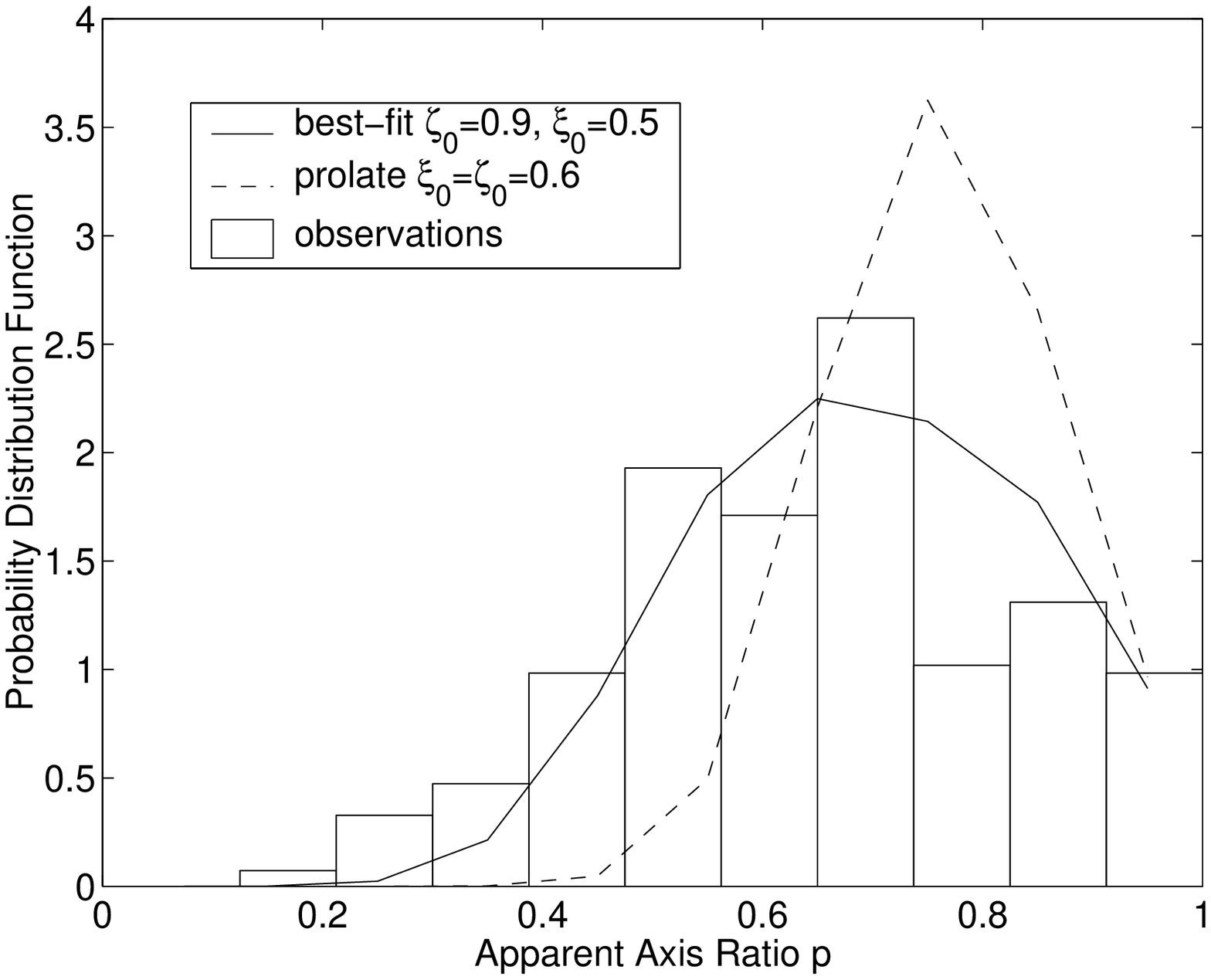}
\caption{A comparison of the observed axis ratios from the ammonia data set
(Jijina et al. 1999) with the best fit assuming triaxial cores (solid line).  
Also shown is the projected distribution for intrinsic distributions with
$\zeta_0 = \xi_0$, i.e., cores which are more prolate in shape (dashed line). 
Both calculated distributions are binned in ten intervals, as is the data set.}
\label{bestfit}
\end{figure}

Finally, Figure~\ref{bestfit} presents a comparison of the best fit of the
projected distribution (based on the $\chi^2$ values) with the Jijina et
al. (1999) data.  
For further comparison, we also display the best fit among 
the projected distributions that arise from intrinsic distributions with 
$\zeta_0 = \xi_0$, i.e., that emphasize prolate shapes.
Notice that such a distribution does not fit the data well;
the distribution is much too strongly peaked.
This strong peaking can
be traced back to the very strongly peaked nature of the observed distribution
from any {\it single} prolate object (see Fig. \ref{binneyobandpro}).

\section{Discussion}

Why are triaxial bodies favored in this analysis? A very important
factor in analyzing the the projected distribution $\phi(p)$ is its shape
near $p=1$. The pronounced drop towards $p=1$ (see
Fig. 4) favors 
triaxial bodies since all such bodies viewed in projection yield $\phi(p) 
\rightarrow 0$ as $p \rightarrow 1$ (see Fig. 2). 
Within the confines
of the axisymmetric assumption, a decrease in $\phi(p)$ towards unity
favors prolate over oblate objects (Ryden 1996) because the former are 
less likely to appear nearly circular in projection (see Fig. 1). However,
once the axisymmetric restriction is dropped, all triaxial clouds can
satisfy the need for a decrease in $\phi(p)$ towards unity. In this case, the
broad peak in the distribution and the fact that a significant number of
cores with $p \approx 1$ are still present favors
the near-oblate triaxial clouds rather than the near-prolate ones. As
exhibited in Figure~\ref{binneyobandpro} and Figure~\ref{bestfit}, prolate clouds (or distributions that 
emphasize prolate clouds) yield projected distributions that are too 
sharply peaked and underestimate the data points near $p=1$.

It is interesting to note that our results are qualitatively similar
to the conclusions of Lambas et al. (1992) and Binney \& Merrified (1998)
that triaxial, but more nearly oblate, shapes are always preferred when 
de-projecting elliptical galaxy shapes. This is despite the different
physical forces at play in the support of galaxies versus cloud cores,
though both represent self-gravitating systems.

While a triaxial shape for dense cores clearly favors non-equilibrium
phenomena influencing their evolution (see discussion in \S\ 1), the
near-oblateness also implies that the cores may not be particularly far
from equilibrium. Oblateness is theoretically consistent with 
models of magnetically and/or rotationally supported equilibria
(e.g., Mouschovias 1976; Tomisaka, Ikeuchi, \& Nakamura 1989).
In contrast, prolate equilibria have proven exceedingly difficult
to construct, requiring exotic effects whose presence has not been
established (see \S\ 1). Near-equilibrium and near-oblate cores are
also consistent with the overwhelming observational evidence for
near virial equilibrium in cores (e.g., Myers \& Goodman 1988),
and the evidence for preferential flattening of cores along the 
direction of the mean magnetic field, shown by the correlation 
$B \propto \rho^{1/2}$ between magnetic field
strength $B$ and density $\rho$ (see Crutcher 1999 and discussion in
Basu 2000). Most generally, this preferential flattening of cores along
{\it one} direction, parallel to the mean magnetic field, favors an
oblate (or near-oblate) geometry, in which one direction is especially 
flattened; in a prolate geometry, two perpendicular directions must have 
the same degree of flattening.

\begin{figure}[h]
\plotone{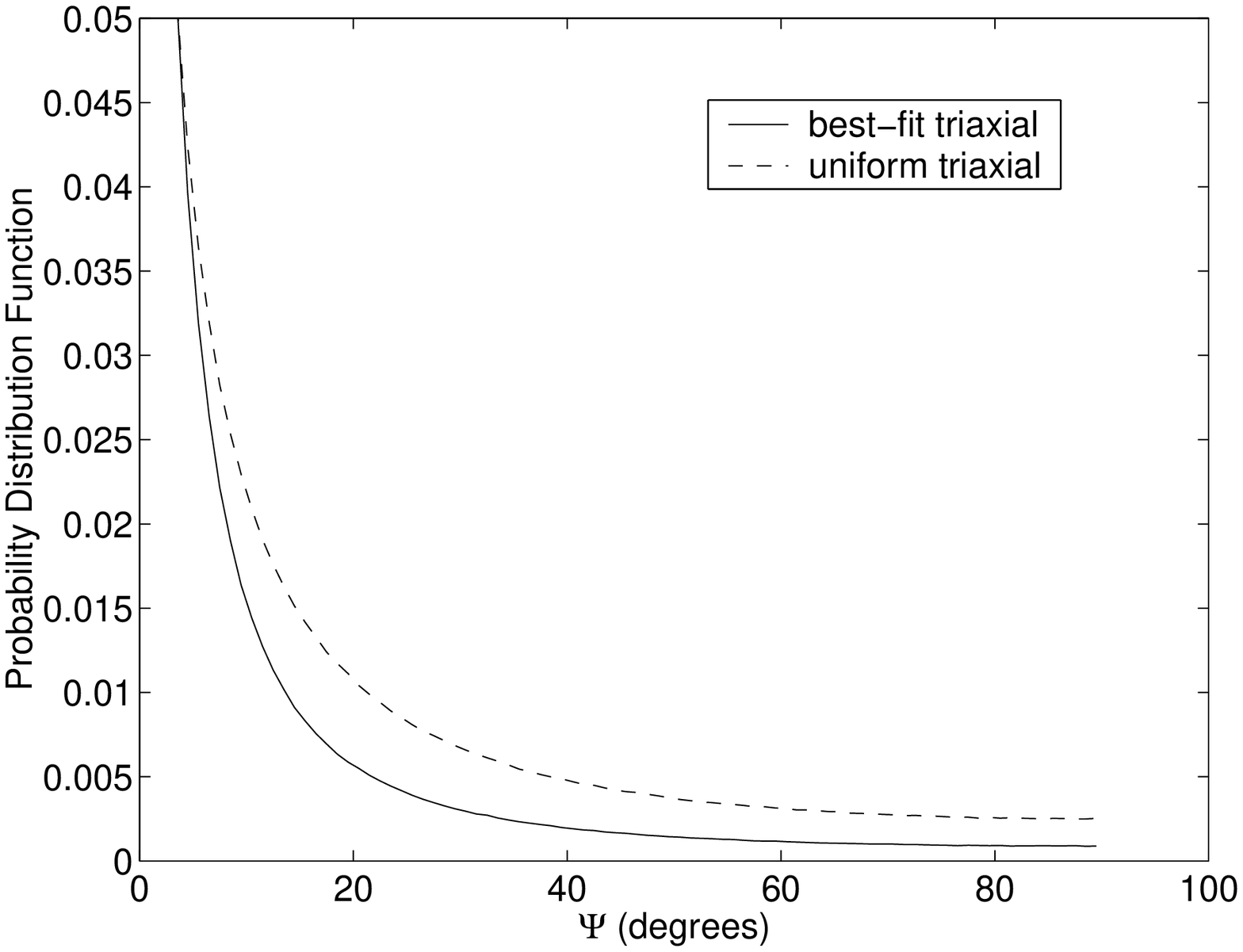}
\caption{Probability density of observing an offset angle $\Psi$
between the projected magnetic field direction and projected minor axis
for a distribution of
triaxial bodies for which the short axis coincides with the mean magnetic field
direction. We use either our best-fit (to the Jijina et al. data set) 
distribution of axis ratios (solid line) or a uniform distribution of 
axis ratios (dashed line).}
\label{poldir}
\end{figure}

If cores are indeed flattened along one preferred direction which corresponds
to that of the mean magnetic field, one may infer the 
distribution of observed offset angles $\Psi$ between the projected magnetic
field direction and the projected minor axis of the core. 
As shown by Basu (2000), a triaxial body seen in projection will
allow a finite probability of viewing any nonzero $\Psi$, although there
will be a bias toward $\Psi=0$. Here, we calculate a
probability distribution for viewing $\Psi$ using our best fit intrinsic
triaxial shape distribution, and assuming the shortest axis of any core 
corresponds to the
mean magnetic field direction. (See Basu 2000 for a discussion of how 
$\Psi$ is calculated.) Figure~\ref{poldir} shows the resulting 
probability distribution
for an intrinsic axis ratio distribution with peaks at 0.5 and 0.9, 
and $\sigma = 0.1$. For comparison, the distribution of $\Psi$ for a 
uniform distribution of triaxial bodies is also shown. There is not a big
difference between the two, although the best fit distribution, which
emphasizes near-oblate clouds, yields a smaller probability for 
observing large $\Psi$ (note that in the oblate limit, a cloud only allows
observations of $\Psi=0$).
These probability distribution functions act as a prediction
for future distributions of $\Psi$ obtained
through submillimeter polarimetry of a large number of dense cores 
(e.g., see Matthews \& Wilson 2000; Ward-Thompson et al.  2000 for a few
recent measurements). The prediction of a bias
toward $\Psi=0$ due to preferential flattening along the mean magnetic field
direction is consistent with the observed $B \propto \rho^{1/2}$ relation and 
contrasts with that expected if turbulent motions dominate magnetic
forces during core formation, in which case there should be no 
correlation of $\Psi$ toward any value.

\section{Summary}

In \S\ 3 we considered the assumption of pure axisymmetric
oblate or prolate cores analytically.  We found that this
assumption led to unphysical results, so we rejected the hypothesis
that cores are axisymmetric.  In \S\ 4, we investigated triaxial
cores statistically.  We found that for the Jijina et al. (1999) data
set, intrinsic mean axis ratios of $0.5$ and $0.9$ best fit
the observations, and for the Lee \& Myers (1999) data set, mean axis
ratios of $0.3$ and $0.9$ best fit the observations,
for assumed Gaussian distributions with standard deviation $\sigma$ of 0.1.
Additionally, choosing $\sigma$ in the range [0.05,0.2] gives essentially 
the same result.

It is worth reiterating
that we find that one of the best fit intrinsic axis ratios is 
always quite a bit larger than the other axis ratio,
which suggests that cores are preferentially flattened in one direction, and
close to oblateness.
While triaxiality implies a non-equilibrium state for most dense cores,
the near-oblate shape implies that the initial conditions for
collapse may not be particularly far from equilibrium.

\begin{acknowledgements}

This research was supported by a grant from NSERC, the Natural Sciences and
Engineering Research Council of Canada. C. E. J. acknowledges
financial support from an NSERC postdoctoral fellowship.

\end{acknowledgements}

\end{document}